\documentclass[12pt,preprint]{aastex}

\newcommand{\teff}{$T_{\rm eff}$} 
\newcommand{\kms}{km s$^{-1}$}
\begin{document}

\title{Rubidium and lead abundances in giant stars of the globular 
clusters M4 and M5\footnote{Based on 
observations made with the Magellan Clay Telescope at Las Campanas 
Observatory.}}

\author{David Yong}
\affil{Research School of Astronomy and Astrophysics, Australian National 
University, Mount Stromlo Observatory, Cotter Road, Weston Creek, ACT 2611, 
Australia}
\email{yong@mso.anu.edu.au}

\author{David L.\ Lambert}
\affil{The W.J. McDonald Observatory, University of Texas, Austin, TX 78712}
\email{dll@astro.as.utexas.edu}

\author{Diane B.\ Paulson}
\affil{NASA's Goddard Space Flight Center, Code 693.0, Greenbelt MD
20771}
\email{diane.b.paulson@gsfc.nasa.gov}

\author{Bruce W.\ Carney}
\affil{Department of Physics \& Astronomy, University of North
Carolina, Chapel Hill, NC 27599-3255}
\email{bruce@physics.unc.edu}

\begin{abstract}

We present measurements of the neutron-capture elements Rb and Pb for
bright giants in the globular clusters M4 
and M5. The clusters are of similar metallicity
([Fe/H] $\simeq -1.2)$ but M4 is decidedly $s$-process enriched 
relative to M5: [Ba/Fe] = +0.6 for M4 but 0.0 for M5. The Rb and Pb abundances 
were derived by comparing synthetic spectra
with high-resolution, high signal-to-noise ratio spectra obtained with 
MIKE on the Magellan telescope. Abundances of Y, Zr, La, and Eu
were also obtained. In M4, the mean abundances from 12 giants are
[Rb/Fe] = 0.39 $\pm$ 0.02 ($\sigma$ = 0.07), 
[Rb/Zr] = 0.17 $\pm$ 0.03 ($\sigma$ = 0.08), and 
[Pb/Fe] = 0.30 $\pm$ 0.02 ($\sigma$ = 0.07). In M5, the mean abundances from two giants are
[Rb/Fe] = 0.00 $\pm$ 0.05 ($\sigma$ = 0.06), 
[Rb/Zr] = 0.08 $\pm$ 0.08 ($\sigma$ = 0.11), and 
[Pb/Fe] = $-$0.35 $\pm$ 0.02 ($\sigma$ = 0.04). 
Within the measurement uncertainties, the abundance ratios [Rb/Fe], 
[Pb/Fe] and [Rb/X] for X = Y, Zr, La
 are constant from star-to-star in each cluster and 
none of these ratios are correlated with O or Na abundances. While M4 has a 
higher Rb abundance than M5, the ratios [Rb/X] are similar in both clusters 
indicating that 
the nature of the $s$-products are very similar for each cluster
but the gas from which M4's stars formed had a higher concentration 
of these products.

\end{abstract}


\section{Introduction}
\label{sec:intro}

Globular clusters continue to provide a source of fascination and frustration 
to both theorists and observers. Two notable accomplishments include the 
use of globular clusters to (a) check the age of the Universe 
(e.g., \citealt{gratton03c}) 
and to (b) test and refine our understanding of stellar 
evolution (e.g., \citealt{renzini88}). 
Despite these successes, globular clusters
continue to present bewildering puzzles. The most persistent puzzle relates 
to chemical composition. 

For many years, globular clusters have been known to exhibit star-to-star
abundance variations for the light elements C, N, O, Na, Mg, and Al 
(e.g., see reviews by \citealt{smith87}, \citealt{kraft94}, 
and \citealt{gratton04}). While 
the amplitude of the star-to-star abundance 
dispersion can vary from cluster to cluster, 
the now familiar anticorrelations between C and N, O and Na, and Mg and Al 
reveal that the abundance variations are likely produced during 
hydrogen burning at high temperatures via the CNO, Ne-Na, and Mg-Al 
cycles. (The O-Na and Mg-Al anticorrelations are not seen in field stars.) 
However, the stars responsible for the nucleosynthesis and the 
nature of the pollution mechanism(s) remain poorly understood (see 
\citealt{lattanzio06} for a recent summary). 

One possible explanation for the observed abundance anomalies 
is internal mixing and nucleosynthesis 
(e.g., \citealt{sm79,charbonnel95}) within the present cluster members, 
the so-called evolutionary scenario.
 The systematic variation of 
the C and N \citep{ss91} and Li \citep{grundahl02} abundances 
with luminosity along the red giant branch demand an evolutionary
component to the star-to-star abundance variations.
Dredge-up of CN-cycled material accounts for the C and N variations. Development
of a giant's convective envelope leading to mixing with highly Li-depleted
gas accounts for the decline of the Li abundance with increasing luminosity.
The proton-capture reactions causing the O, Na, Mg, and Al variations demand
 much higher temperatures and much deeper mixing than those required for
CN-cycling. Such 
mixing is not predicted by standard theoretical models of red giants 
and the discovery of the O, Na, Mg, and Al anomalies in main sequence 
stars (e.g., \citealt{briley96}, \citealt{gratton01}) eliminates
deep mixing as a viable explanation for the O-Al variations. The 
interiors of main sequence stars are too cool to process Ne to Na or Mg to Al.
Therefore, the cluster gas must have been inhomogeneous when the present 
stars were formed. This alternative explanation for the abundance anomalies 
is the so-called primordial scenario. 

In the primordial scenario, 
intermediate-mass asymptotic giant branch stars (IM-AGBs) from the generation
to which the observed stars belong have long been 
considered candidates for synthesizing the abundance variations 
\citep{cottrell81}. In IM-AGBs, the convective envelope can reach the top 
of the hydrogen-burning shell, a process called hot-bottom burning. For 
sufficiently massive and metal-poor AGBs, the temperatures at the
base of the convective envelope can exceed 
100 million degrees thereby allowing the efficient operation of the CNO, 
Ne-Na, and Mg-Al cycles (e.g., \citealt{karakas03}). 
That IM-AGBs do not alter the abundances of the 
alpha or iron-peak elements (as required by observations) 
adds to their qualitative appeal. However, quantitative tests reveal 
problems with the IM-AGB primordial scenario. Theoretical yields
from IM-AGBs combined with a chemical evolution model \citep{fenner04} 
suggest that O is not sufficiently depleted, Na is overproduced, Mg is 
produced rather than destroyed, the isotope ratios of Mg do not match the 
observations, and the sum of C+N+O increases substantially in contrast to 
the observations. 
\citet{ventura05a,ventura05b,ventura05c} find that many of the flaws noted 
above can be alleviated when IM-AGB yields are calculated using a revised 
treatment for convection and mass-loss. However, \citeauthor{ventura05a} 
note that problems persist, namely with the Mg isotope ratios, and warn 
that the predictive power of the current AGB models is limited. Recently, 
\citet{prantzos06}, \citet{smith06}, and \citet{decressin06}
suggest that the winds from massive stars may be more 
promising candidates than IM-AGBs. There is no satisfactory explanation for 
the complex patterns for the light element abundances exhibited by every
well studied Galactic globular cluster. Therefore, our present 
understanding of globular cluster chemical evolution and/or stellar 
nucleosynthesis is incomplete. 

Determinations of the stellar abundances of the trans-iron or heavy elements 
offer clues to the history behind the chemical evolution of
globular clusters. Here, we provide novel information -- the Rb and Pb abundances --
for giants in M4 and M5, a pair of clusters of similar metallicity but
with distinctly different levels of $s$-process products.
The quintessential $r$-process element Eu has similar abundances in the
two clusters and, indeed, across 
the collection of Galactic globular
clusters. In sharp contrast, the $s$-process products are more evident in M4 than in
M5 and other clusters of similar metallicity: [Ba/Fe] is about +0.6 in M4 but
0.0 in M5. The questions - Are there differences in the Rb and Pb abundances 
between this pair of clusters? and Are the star-to-star variations in the
abundances of light elements (O, Na, Mg, and Al) reflected in variations among the
abundances of Rb and Pb? -- seem likely to probe the origins of the
$s$- and $r$-process products for globular clusters.

Due to a critical 
branching point in the $s$-process path at $^{85}$Kr, the abundance 
of Rb relative to Sr, Y, or Zr can differ by a factor of 10 depending upon 
the neutron density at the $s$-process site. In the case of AGB stars, the
neutron density in the He-shell is dependent on the stellar mass 
(e.g., see 
\citealt{tomkin83}, \citealt{lambert95}, \citealt{busso99}, and 
\citealt{abia01} for further details). Since the isotopes of Pb and Bi are 
the last stable nuclei on the $s$-process path, the $s$-process terminates at these
elements and overabundances of Pb and Bi will arise if seed nuclei are
shuffled by neutron captures down the entire $s$-process path.
In particular, metal-poor AGB stars may 
produce large overabundances of Pb and Bi if the neutron supply per seed 
exceeds a critical value (e.g., see \citealt{goriely01}, 
\citealt{travaglio01}, and \citealt{busso01} for further details). 
The suspicion is that the star-to-star abundance variations for light
elements are due to contamination by IM-AGBs.
Some contend that IM-AGBs also synthesize $s$-process nuclides
and then one might expect to see star-to-star variations in the
Rb and Pb abundances as well as correlations with light element abundances. 

To further examine the possible role of IM-AGBs in the chemical evolution 
of globular clusters, \citet{rbpbsubaru} 
measured Rb and Pb in NGC 6752 and M13, the two 
clusters that exhibit the largest amplitude for Al variations. It was found that
 the abundance ratios [Rb/Zr] and [Pb/Fe] were constant from 
star-to-star within the measurement uncertainties.
 If IM-AGBs do synthesize
Rb and Pb, then they may not be responsible for the abundance variations. 
On the other hand, if IM-AGBs are responsible for the abundance variations, 
they cannot synthesize Rb or Pb. 

In this paper, we extend the measurements of Rb and Pb to the globular 
clusters M4 and M5. While these clusters are more metal-rich than NGC 6752 
or M13, both M4 and M5 are known to exhibit large dispersions 
and correlations for the light element abundances [see pioneering 
studies on CN bimodality by \citet{norris81a} and \citet{smith83} as well 
as recent high-resolution spectroscopic studies by \citet{M4,M5}, 
\citet{ramirez03}, and references therein]. In particular, as noted above,
M4 is remarkably, perhaps uniquely among globular clusters, enriched in $s$-process
products. 

\section{Observations and data reduction}
\label{sec:data}

The targets included 12 stars in M4 previously studied by \citet{M4} 
and two stars in M5 previously studied by \citet{M5} and \citet{ramirez03}.
The main focus was to observe a large number of stars in M4 and though we 
were restricted to the brightest giants, we note that the sample spans 
a considerable range of the known star-to-star elemental abundance variations. 
The smaller sample in M5 is due to the fact that those observations were 
conducted during other observing programs when primary targets were 
unavailable. In Table \ref{tab:param}, we list the program stars. 
The stellar identifications for M4 and M5 are from \citet{lee77} and 
\citet{arp62} respectively. 

The observations were performed with the Magellan
Telescope using the Magellan Inamori Kyocera 
Echelle spectrograph (MIKE; \citealt{mike}) on 2004 June 12-13, July 16, 
and July 18. A 0.35\arcsec\ 
slit was used providing a resolving power of 
R $\equiv \lambda/\Delta\lambda$ =
55,000 in the red and R $=$ 65,000 in 
the blue per 4 pixel resolution element with wavelength coverage from
3800~\AA~to 8500~\AA. The IRAF\footnote{IRAF (Image Reduction and Analysis
Facility) is distributed by the National Optical Astronomy
Observatory, which is operated by the Association of Universities
for Research in Astronomy, Inc., under contract with the National
Science Foundation.} package of programs 
was used for most of the data reduction. In order 
to correct for the fact that the lines are severely tilted with respect 
to the orders and the tilt varies across the CCD, we used the 
{\sc mtools}\footnote{http://www.lco.cl/lco/magellan/instruments/MIKE/reductions/mtools.html} 
set of tasks written by Jack Baldwin to extract the spectral orders. 
In the one-dimensional wavelength-calibrated normalized spectra, the 
typical signal-to-noise ratio was 70 per pixel at 4050~\AA~(140 per
resolution element) and 400 per pixel at 7800~\AA~(800 per resolution 
element). 

\section{Analysis}
\label{sec:analysis}

\subsection{Stellar parameters and the iron abundance}
\label{sec:param}

The required stellar parameters for an abundance analysis are the 
effective temperature (\teff), the surface gravity ($\log g$), and the 
microturbulent velocity ($\xi_t$). 
Our analysis techniques closely follow \citet{rbpbsubaru} in which 
we determined these values adopting a 
traditional spectroscopic approach. Using routines in IRAF, we measured the 
equivalent widths (EWs) for a set of Fe\,{\sc i} and Fe\,{\sc ii} lines. 
The set of Fe lines was identical to those used by \citet{mghsubaru} for 
which the $gf$-values were taken from \citet{lambert96}, \citet{fe2}, 
\citet{blackwell95} and references therein. Model atmospheres were taken 
from the \citet{kurucz93} local thermodynamic equilibrium (LTE) stellar
atmosphere grid and we interpolated within the grid when necessary. We 
used the LTE stellar line analysis program {\sc moog} \citep{moog} to determine 
abundances for a given line. We adjusted \teff~until there was no trend 
between the abundances from Fe\,{\sc i} lines 
and the lower excitation potential. 
We adjusted $\log g$ until the abundances from Fe\,{\sc i} and Fe\,{\sc ii} 
lines were in agreement. We adjusted $\xi_t$ until there was no trend between
the abundances from Fe\,{\sc i} lines and EW. This process was iterated 
until 
all three parameters were simultaneously constrained. The final [Fe/H] was
the mean from all Fe lines assuming a solar abundance 
log~$\epsilon$(Fe) = 7.50. The stellar parameters are given in Table 
\ref{tab:param} and we estimate the internal errors to be \teff~$\pm$~50~K, 
$\log g~\pm$~0.2 dex, and $\xi_t~\pm$~0.2 \kms. 

As an additional check on our surface gravities (and analysis techniques), 
we compared our 
$\log g$ values with the $Y^2$ isochrones \citep{y2}. We adopted the 
13 Gyr isochrones with [$\alpha$/Fe] = +0.3, our 
spectroscopic \teff, and interpolated between the two closest metallicities
z = 0.004 and z = 0.001. We found that our surface gravities were 
in fair agreement with the $Y^2$ isochrones, 
$\log g_{\rm spec} - \log g_{\rm isochrone} = -0.22$ $\pm$ 0.07 
($\sigma$ = 0.26). 

The stellar parameters are in good agreement with those obtained by 
\citet{M4,M5} and \citet{ramirez03}. For the 12 M4 giants, the mean 
differences (This Study $-$ \citet{M4}) are 
$\Delta$\teff~= 71 $\pm$ 17 ($\sigma$ = 58 K), 
$\Delta\log g$ = 0.02 $\pm$ 0.05 ($\sigma$ = 0.17), 
$\Delta\xi_t$ = 0.04 $\pm$ 0.04 ($\sigma$ = 0.15 \kms), and 
$\Delta$[Fe/H] = $-$0.05 $\pm$ 0.01 ($\sigma$ = 0.05). For the 2 M5 giants, 
the mean differences (This Study $-$ \citet{ramirez03}) are 
$\Delta$\teff~= 20 $\pm$ 5 ($\sigma$ = 7 K), 
$\Delta\log g$ = $-$0.15 $\pm$ 0.15 ($\sigma$ = 0.21), 
$\Delta\xi_t$ = $-$0.16 $\pm$ 0.03 ($\sigma$ = 0.04 \kms), and 
$\Delta$[Fe/H] = 0.05 $\pm$ 0.02 ($\sigma$ = 0.03). For M5 IV-81, our 
\teff~is 105 K higher, $\log g$ is 0.3 dex higher, $\xi_t$ is 
0.02 \kms~lower, and [Fe/H] is 0.12 dex higher than the values determined 
by \citet{M5}. 

\subsection{Rubidium abundances}

\begin{figure}
\epsscale{0.8}
\plotone{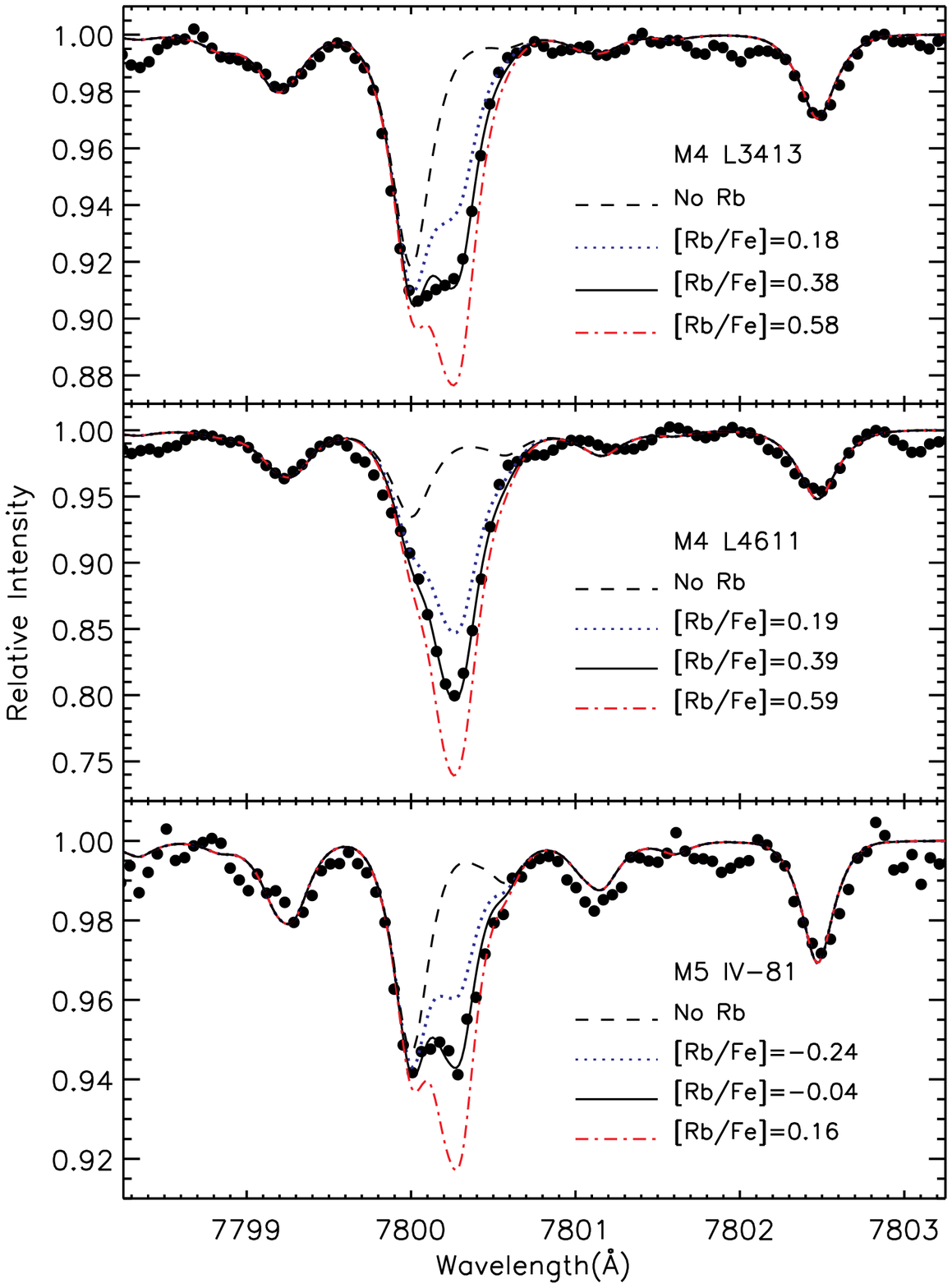}
\caption{Observed spectra (filled circles) for M4 L3413 (upper panel), 
M4 L4611 (middle panel), and M5 IV-81 (lower panel) near the 7800~\AA~Rb 
feature. Synthetic spectra with different Rb abundances are shown 
(the solid line represents the best fit).\label{fig:rbfit}}
\end{figure}

The abundances for Rb were determined via spectrum synthesis of the 
7800~\AA~Rb\,{\sc i} line (see Figure \ref{fig:rbfit}). 
For all stars in our sample, this Rb line is blended with a 
Si\,{\sc i} line as well as with weak CN lines and therefore 
an equivalent width analysis 
is not possible. Even in the coolest stars, the 
Rb line is only 10\% deep relative to the continuum such that accurate 
abundances can be derived only from high-resolution, high 
signal-to-noise ratio spectra. The adopted wavelengths and relative strengths 
for the isotopic and hyperfine-structure components of Rb were identical to 
\citet{lambert76} and \citet{tomkin99}. We assumed a solar isotope ratio 
$^{85}$Rb/$^{87}$Rb = 3 and the macroturbulent broadening was fixed by 
fitting the profile of the nearby 7798~\AA~Ni\,{\sc i} line. Synthetic
spectra were generated using {\sc Moog} and the Si and Rb abundances 
were varied to obtain the best fit to the observed spectrum. 
Our tests confirmed the finding by \citet{lambert76} that Rb isotope ratios 
cannot be measured from the 7800~\AA~line due to the hyperfine structure
and the small isotopic shift. Not surprisingly, our tests showed that 
the derived Rb abundances are insensitive to the assumed isotope ratio. 
In \citet{rbpbsubaru}, we took the \citet{kurucz84}
solar atlas and measured an abundance log $\epsilon$(Rb) = 2.58 using a model
atmosphere with \teff/$\log g$/$\xi_t$ = 5770/4.44/0.85. Our solar Rb 
abundance is in excellent agreement with the \citet{grevesse98} value, 
log $\epsilon$(Rb) = 2.60. 
The weaker Rb\,{\sc i} resonance line near 7947~\AA~is detected but in a region
affected by 
unidentified blends, atmospheric absorption, and fringing. 
While preliminary 
analyses suggest that the abundances derived from the 7947~\AA~line are 
similar to those derived from the 7800~\AA~line 
(see \citealt{rbpbsubaru} for a more detailed comparison), 
we restrict our analyses to the 7800~\AA~line. 

In this study, we included CN molecular lines when fitting the 7800~\AA~Rb 
line. The CN lines were taken from \citet{kurucz06}. While the inclusion of 
these lines resulted in an improved fit to the local continuum, we note that 
the Rb abundances were unaffected. Even for the coolest stars in M4 for 
which the CN line strength was at a maximum, the 
Rb abundances increased by $\le$ 0.02 dex if the CN lines were omitted. 
For the more metal-poor cluster NGC 6752, we re-analyzed a subset of the 
\citet{rbpbsubaru} spectra and found that the Rb abundances were 
unchanged when CN lines were included in the analysis. 

\subsection{Lead abundances}

\begin{figure}
\epsscale{0.8}
\plotone{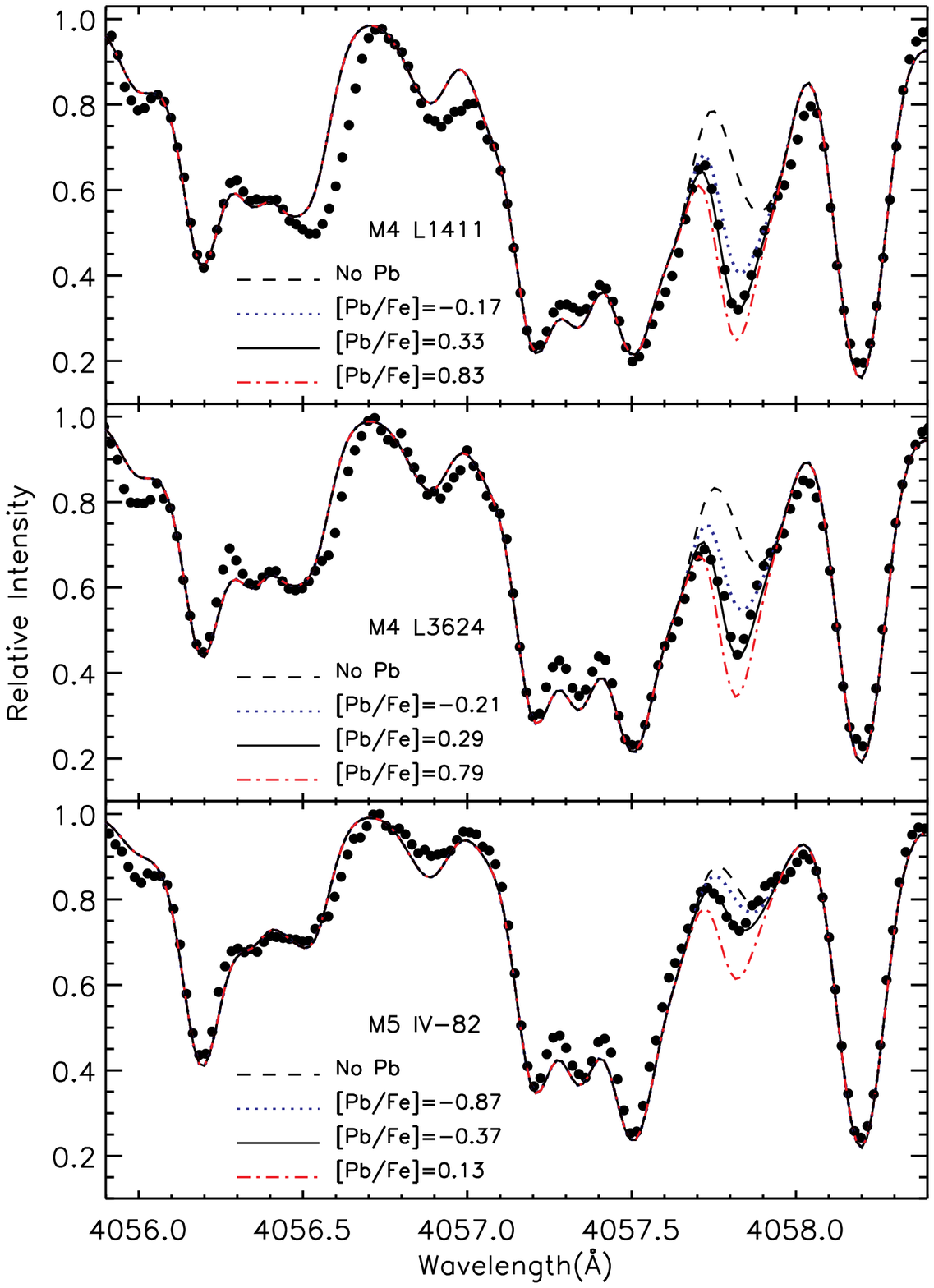}
\caption{Observed spectra (filled circles) for M4 L1411 (upper panel), 
M4 L3624 (middle panel), and M5 IV-82 (lower panel) near the 4058~\AA~Pb 
feature. Synthetic spectra with different Pb abundances are shown 
(the solid line represents the best fit).\label{fig:pbfit}}
\end{figure}

The abundances for Pb were determined via spectrum synthesis of the 
4058~\AA~Pb\,{\sc i} line (see Figure \ref{fig:pbfit}).
 High-resolution, high signal-to-noise ratio 
spectra are essential for measuring Pb abundances 
because the region near 4058~\AA~is crowded with 
molecular lines of CH as well as Mg, Ti, Mn, Fe, and Co atomic lines. 
Since M4 and M5 are more metal-rich than NGC 6752 and M13, the 
4058~\AA~region is more crowded and the uncertainties in the derived 
Pb abundances are greater. However, we note that our syntheses 
provide a very good fit to the observed 
spectra. The macroturbulent broadening was fixed by fitting the profiles 
of nearby lines. The $gf$-value of the Pb\,{\sc i} line 
was the same as that used by \citet{aoki02} 
as were the hyperfine-structure and isotopic components. We assumed 
a solar isotope ratio and our tests confirmed that the derived Pb abundances 
do not depend upon the assumed isotope ratio. For the solar Pb
abundance, we adopted log~$\epsilon$(Pb) = 1.95 from \citet{grevesse98}. 

\subsection{Additional elements}

The abundances for O, Na, Mg, Al, Y, Zr, La, and Eu were also measured 
in the program stars using the same lines as \citet{67522}. Y and Zr were 
selected since the 
ratios [Rb/Y] and [Rb/Zr] are 
sensitive to the neutron density at the site of the 
$s$-process. O, Na, Mg, and Al were 
measured since they are known to vary from star-to-star in these and other 
globular clusters. Re-measuring these 
elements in M4 and M5 provides a check to see if our abundance 
determinations are on the same scale as other investigators. Similarly, the 
neutron-capture elements La and Eu were measured (hyperfine and/or isotopic 
splitting was included). 
When deriving O abundances, we generated synthetic spectra to account 
for possible blending from CN molecular lines. We assumed C and N 
abundances interpolated as a function of Na with the C and N abundances 
taken from \citet{M4,M5} and \citet{smith97}. A subset of abundance 
determinations were conducted using spectrum synthesis to account for 
possible blends. 
In Table \ref{tab:abund} we present the measured elemental abundances for 
O, Na, Mg, Al, Rb, Y, Zr, La, Eu, and Pb. The adopted solar abundances 
were 8.69, 6.33, 7.58, 6.47, 2.60, 2.24, 2.60, 1.13, 0.52, and 1.95
respectively. These values are identical to those we used in previous papers 
and were taken from \citet{grevesse98} for all elements 
except O \citep{allendeO}, La \citep{la}, and Eu \citep{eu}. 
Table \ref{tab:abund} also contains the Zr abundances shifted 
onto the \citet{smith00} scale.

For M4, the measured elemental abundances 
are in good agreement with \citet{M4}. 
The mean differences (This Study $-$ \citet{M4}) are 
$\Delta$[O/Fe] = 0.34 $\pm$ 0.03 ($\sigma$ = 0.11), 
$\Delta$[Na/Fe] = 0.14 $\pm$ 0.03 ($\sigma$ = 0.12),
$\Delta$[Mg/Fe] = 0.14 $\pm$ 0.02 ($\sigma$ = 0.06),
$\Delta$[Al/Fe] = 0.05 $\pm$ 0.03 ($\sigma$ = 0.10),
$\Delta$[La/Fe] = 0.01 $\pm$ 0.03 ($\sigma$ = 0.10), and 
$\Delta$[Eu/Fe] = 0.03 $\pm$ 0.02 ($\sigma$ = 0.07). The agreement for 
Al, La, and Eu is excellent. 
The largest discrepancy is for O and differences in the 
adopted solar abundances can account for 0.24 dex leaving a 0.10 dex  
residual. While both studies use the same O lines and $gf$ values, they 
employ spectrum synthesis which presumably involves a different set of 
atomic and molecular lines. 
A small offset also exists for 
Na and Mg which could be due to the different set of lines employed. 

For M5, the elemental abundances are in fair agreement with \citet{ramirez03}. 
The mean differences (This Study $-$ \citet{ramirez03}) are 
$\Delta$[O/Fe] = 0.27 $\pm$ 0.14 ($\sigma$ = 0.20), 
$\Delta$[Na/Fe] = $-$0.05 $\pm$ 0.10 ($\sigma$ = 0.14),
$\Delta$[Mg/Fe] = 0.11 $\pm$ 0.06 ($\sigma$ = 0.08),
$\Delta$[Zr/Fe] = $-$0.05 $\pm$ 0.19 ($\sigma$ = 0.27),
$\Delta$[La/Fe] = $-$0.03 $\pm$ 0.03 ($\sigma$ = 0.04), and 
$\Delta$[Eu/Fe] = $-$0.06 $\pm$ 0.09 ($\sigma$ = 0.13). For O, the difference 
can be attributed to the adopted solar abundance. For the single star also 
studied by \citet{M5}, the measured abundances are in good agreement 
$\Delta$[X/Fe] $\le$ 0.07 dex. The difference for La is 0.14 dex (possibly 
due to the different set of lines) and 0.58 dex for O.
The adopted solar abundances can account for 0.24 dex leaving 
a 0.34 dex residual. Dr I.\ Ivans kindly sent the synthetic and 
observed spectra near the 6300 \AA~and 6363 \AA~used to derive
O abundances. 
The difference in the measured O abundances between the two studies 
is $\Delta\log~\epsilon$(O) = 0.14 dex. The difference in the measured 
Fe abundance is 0.12 dex. The remaining 0.08 dex residual may be 
attributed to differences in the spectrum synthesis analysis. 

The adopted line lists and measured equivalent widths are 
presented in Table \ref{tab:line}. In Table \ref{tab:parvar}, the abundance 
dependences on the model parameters are given. 

\section{Discussion}
\label{sec:discussion}

\subsection{The heavy-element canvas}

The heavy elements are synthesized by neutron-captures in the
$s$- and the $r$-process. (We ignore here the nuclides, all of
low abundance, referred to as $p$-nuclides.) In all but material
dominated by
$s$-processed products, Eu is a signature
element for the $r$-process. There is evidence from analyses of
field stars, especially those severely enriched in $r$-process
products that the relative abundances for the $r$-process of
nuclides from Ba to Eu are invariant with metallicity 
(e.g., see \citealt{cowan06}).
This invariance does not extend to the ratio of heavy products (i.e., Eu) 
to light products (i.e., Sr, Y, and Zr).
 In a solar mix of elements, Ba is 
taken as a measure of the $s$-process with but a slight contamination from
the $r$-process \citep{burris00}. The [Ba/Eu] ratio of a star is widely
taken to indicate the relative mix of $s$- to $r$-processed
material.

In the case of ratios such as [Eu/Fe] and [Ba/Fe] 
 for globular cluster stars, three questions
arise: Is there an intercluster spread in the ratios? Is there an
intra-cluster spread, especially for clusters exhibiting a large
spread among light element abundances? How do the cluster ratios
compare with those found for the field stars, as a function of [Fe/H]?

\subsubsection{The [Eu/Fe] ratio}

With but the single known exception of Ruprecht 106, the [Eu/Fe]
ratio across
 the collection of examined Galactic globular
clusters with [Fe/H] from about $-$1.0 to $-2.5$ appears to be
single-valued.
Values of [Eu/Fe] in the literature for well-sampled
clusters with [Fe/H] $< -1$ span a reported range of only 0.2 dex
from about $+0.35$ to $+0.55$ with no obvious trend with [Fe/H].
[Ruprecht 106 is an outlier at $0.0$
from analyses of two stars \citep{brown97}. M80 is probably 
also an outlier at +1.0, but the analyses were based on less 
than ideal spectra, R=18,000 \citep{csp04}.] 
This range is gathered from examination of the
extensive literature; references cited by \citet{gratton04} 
were read and supplemented by more recent papers. No
attempt has been made by us to correct published values to
a common standard. It would be premature to claim that the
0.2 dex spread in [Eu/Fe] is real and thus not entirely attributable to
cumulative errors of measurement. Our results
[Eu/Fe] = $+0.41$ for M4 and $+0.53$ for M5 fall within this
range and, as noted above, are consistent with previously published
results. 
 
Although our samples are small (see also Yong et al. 2006), the stars
chosen for their contrasting light element abundances show no
evidence of a dependence on these light element abundances. This
decoupling of light element abundances from that of Eu is confirmed
by previous studies of well-sampled clusters.
Several authors have compared [Eu/Fe] for globular clusters with
results for Galactic field stars. The evidence is that the [Eu/Fe]
-- [Fe/H] trends (i.e., the apparent independence of [Eu/Fe] on
 [Fe/H] for [Fe/H] $< -0.7$)
are the same for both samples (see, for example, \citet{gratton04},
\citet{james04b}, and \citet{pritzl05}). 

In summary, globular cluster and field stars appear with remarkably few
exceptions and to within measurement errors to have the same
[Eu/Fe] ratio. Our measurements of this ratio for M4 and M5 are
consistent with previous measurements for these clusters and the
common cluster-field star set of data.

\subsubsection{The [Y/Fe], [Zr/Fe], [Ba/Fe], and [La/Fe] ratios}

Among the few heavy elements in abundance analyses
of globular cluster stars from which the $s$-process contribution
may be assessed, Ba is the most widely reported. 
Other elements for which abundance data are available
for several clusters include Y, Zr, and La.

For most globular clusters, the [Ba/Fe] ratio is positive 
([Ba/Fe] $\simeq +0.2$) and independent of [Fe/H] over the interval
$-0.7$ to about $-1.5$. For the few investigated clusters with
[Fe/H] $\le -1.8$, [Ba/Fe] declines to roughly 0.0 at 
[Fe/H] of $-2.0$ to $-2.5$. This run with [Fe/H] for the clusters
is matched by the field stars \citep{james04b}. The spread in
[Ba/Fe] at a given [Fe/H] is about $\pm0.2$ dex, a value
consistent with the errors of measurement, with several outliers.
The sole outlier with [Ba/Fe] above the mean trend is M4 with
an excess of about 0.4 dex in [Ba/Fe]. The outliers below the
mean trend are NGC 3201, Ruprecht 106, 
and Palomar 12 \citep{brown97,gonzalez98} as well as 
NGC 5694 \citep{lee06}.
Data for [La/Fe] is less extensive but suggests [La/Fe] $\simeq +0.1$
with M4 again as an outlier ([La/Fe] $= +0.45$) but with the Ba-outliers
NGC 3201 and Pal 12 as conformers to the mean value. NGC 5694 is again 
an outlier with [La/Fe] = $-$0.26 based on analysis of a single star 
\citep{lee06}. The [Y/Fe] ratio is 
close to zero \citep{james04b}. Our result for M5 at +0.2 from just two
stars may be an (unlikely) outlier. At [Y/Fe] = +0.7, M4 is certainly
an outlier. 

The discussed data, as well as less extensive data on other
heavy elements, show that the vast majority of the
clusters are not
distinguishable from field stars by their abundances of Ba and other heavy
elements. For a given cluster, there is no compelling evidence for a
star-to-star variation correlated with the variations seen among
the light elements for many clusters. There are, however, 
examples of individual cluster stars with $s$-process enrichments,
e.g., the CH stars in M2 and M55 \citep{smith90,briley93}, the 
possible CH star in M22 \citep{mcclure77,vanture92}, 
a Y-rich star in M13 \citep{cohen05}, and a hint of Zr abundance variations 
in 47 Tuc \citep{wylie06}, M5 \citep{ramirez03}, and NGC 6752 \citep{67522}. 
While M15 exhibits 
variations of Ba and Eu, these variations are not correlated with the light 
element abundances and the ratio [Ba/Eu] is constant from 
star-to-star \citep{sneden97,otsuki06}. 

\subsection{Rubidium}

In M4 the mean Rb abundance is [Rb/Fe] = 0.39 $\pm$ 0.02 ($\sigma$ = 0.07) 
and in M5 the mean abundance is [Rb/Fe] = 0.00 $\pm$ 0.05 ($\sigma$ = 0.06). 
As expected, M4 has a higher abundance of Rb 
than M5. While 50\% of the solar Rb abundance can be attributed to the 
$r$-process \citep{burris00}, 
M4 has a slightly lower [Eu/Fe] ratio 
than does M5 and therefore, it would 
be difficult to attribute any Rb excess in M4 relative to M5 as 
being due to a larger $r$-process contribution.
Neither cluster shows any evidence for a dispersion in Rb abundances. That is, 
the scatter in the Rb abundances within M4 and M5 is small and can be 
attributed entirely to the measurement uncertainties. 
The Rb abundances are not correlated with O or Na 
(the two elements that exhibit large star-to-star abundance variations). 

\begin{figure}
\epsscale{0.8}
\plotone{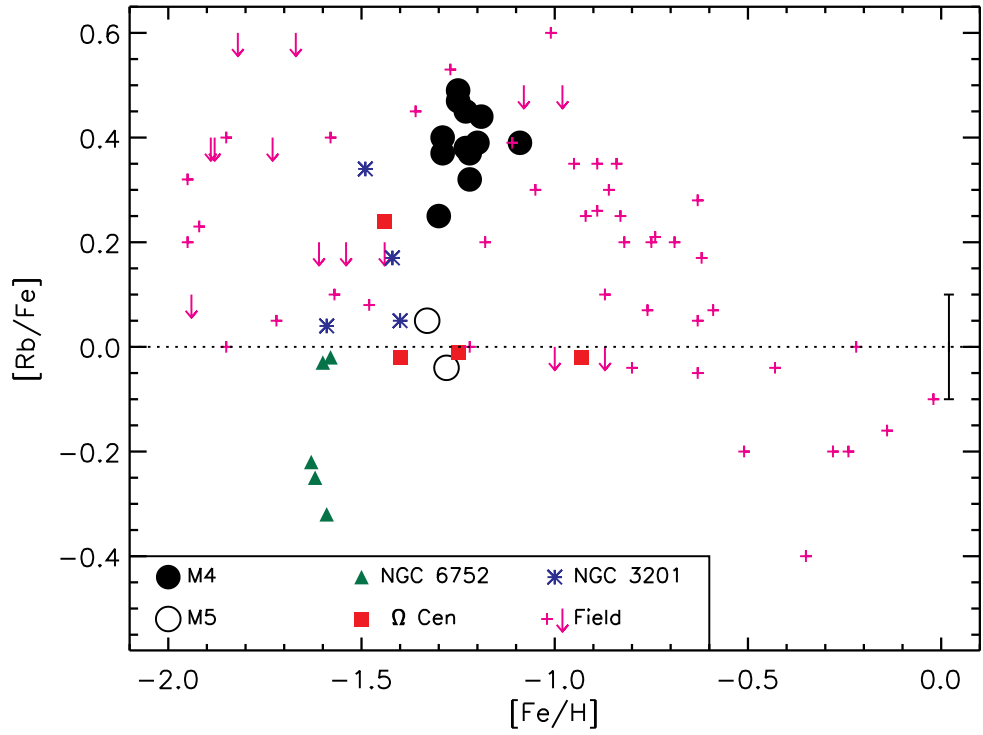}
\caption{[Rb/Fe] vs.\ [Fe/H] for M4 (filled circles) and M5 (open circles). 
Green triangles show NGC 6752 \citep{rbpbsubaru}, red 
squares represent $\omega$ Cen \citep{smith00}, blue asterisks are NGC 3201 
\citep{gonzalez98}, and magenta plus signs and upper limits 
represent field stars from 
\citet{tomkin99} and \citet{gratton94}. A representative error 
bar is shown. The abundances have been shifted onto the \citet{smith00} 
scale.\label{fig:rb}}
\end{figure}

The Rb abundances of M4 and M5 are compared with the limited
data in the literature for other clusters and field stars.
In Figure \ref{fig:rb}, we compare the [Rb/Fe] abundance ratios with 
field dwarfs and giants \citep{gratton94,tomkin99} as well as the 
globular clusters NGC 3201 \citep{gonzalez98}, $\omega$ Cen \citep{smith00}, 
and NGC 6752 \citep{rbpbsubaru}. M5 has [Rb/Fe] ratios that are essentially 
identical to $\omega$ Cen giants at the same metallicity. [We note that 
$\omega$ Cen is now regarded as the nucleus of an accreted dwarf 
spheroidal galaxy \citep{smith00}.] These Rb abundances 
are comparable to the lowest values seen in field stars at the same 
metallicity. 
Given that M4 is $s$-process enriched relative to 
other globular clusters and field stars, it is puzzling that at the 
metallicity of M4, the majority of field stars share the same [Rb/Fe] ratio. 
With the addition of M4 and M5, 
the [Rb/Fe] ratios in NGC 6752 appear rather low compared to other globular 
clusters and field stars. While the sample sizes remain small and the 
analyses have been performed by different investigators, the globular clusters
cover a wider range of [Rb/Fe] values than do the field stars in the same 
metallicity regime. 

\begin{figure}
\epsscale{0.8}
\plotone{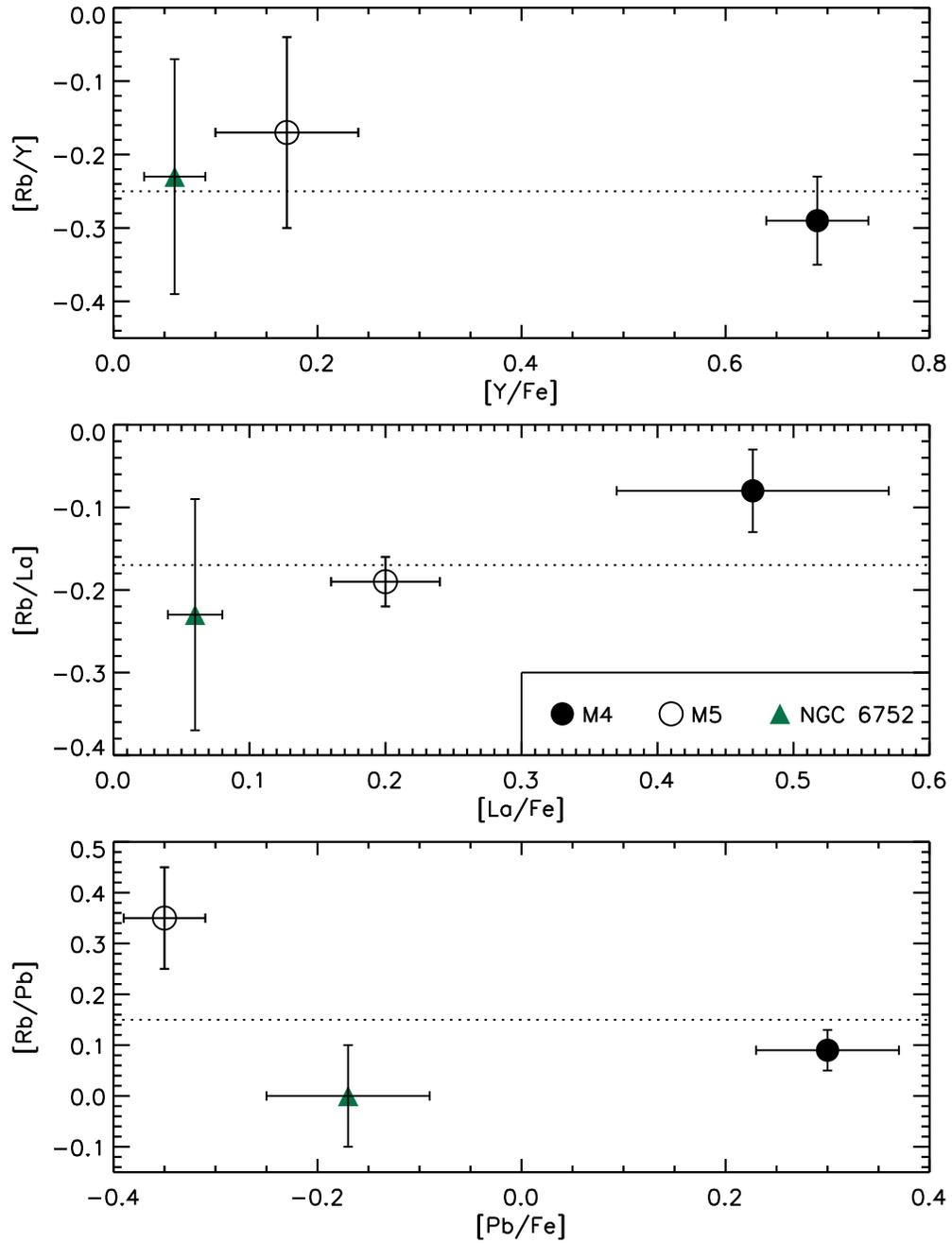}
\caption{[Rb/Y] vs.\ [Y/Fe] (upper), [Rb/La] vs.\ [La/Fe] (middle), and 
[Rb/Pb] vs.\ [Pb/Fe] (lower) for M4, M5, and NGC 6752. 
In each panel, the dotted line represents the mean value. 
\label{fig:new3}}
\end{figure}

\begin{figure}
\epsscale{0.8}
\plotone{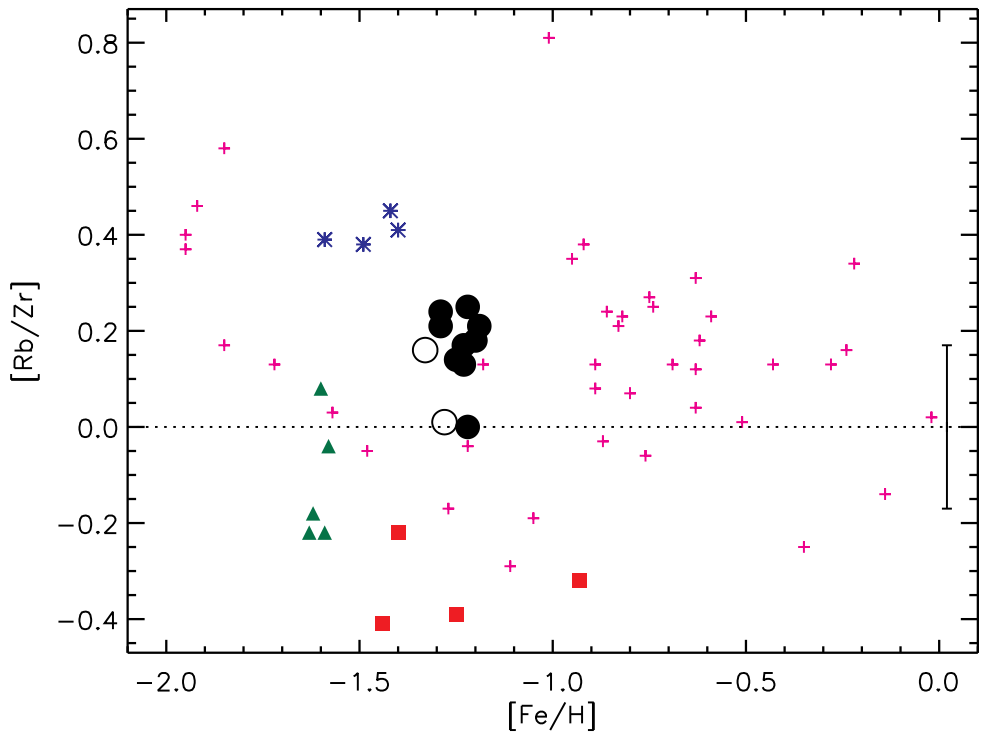}
\caption{[Rb/Zr] vs.\ [Fe/H]. The symbols are the same as in Figure 
\ref{fig:rb}. The abundances have been shifted onto the \citet{smith00} 
scale. A representative error bar is shown.\label{fig:rbzr}}
\end{figure}

The relative abundances of Rb, Y, and Zr are very similar for M4 and M5.
In Figure \ref{fig:new3}, the upper panel shows [Rb/Y] versus [Y/Fe] for
mean values for M4, M5, and NGC 6752. The dotted line corresponds to
[Rb/Y]$ = -0.25$, the mean value for the three clusters. 
Published results for NGC 3201 and $\omega$ Cen do not fall on 
the line but straddle it. This may indicate zero-point differences between our
and other analyses or a real cluster-to-cluster difference among the
Rb, Y, and Zr abundances.
Considering the individual measurements, we find
[Rb/Y] = $-$0.29 $\pm$ 0.02 ($\sigma$ = 0.06) for M4 and 
[Rb/Y] = $-$0.17 $\pm$ 0.10 ($\sigma$ = 0.13) for M5.
For Zr, 
we find [Rb/Zr] = 0.17 $\pm$ 0.03 ($\sigma$ = 0.08) for M4 and 
[Rb/Zr] = 0.08 $\pm$ 0.08 ($\sigma$ = 0.11) for M5.
 (These Zr abundances have 
been shifted by +0.3 dex onto the \citealt{smith00} scale, 
see \citealt{rbpbsubaru} for details.)
Neither the [Rb/Y] nor the
[Rb/Zr] ratio exhibits a 
star-to-star dispersion and neither is 
correlated with O or Na abundances. 
In Figure \ref{fig:rbzr}, we compare the [Rb/Zr] abundance ratio with 
field dwarfs and giants as well as globular cluster giants. Compared 
to the $\omega$ Cen giants at a similar metallicity, M4 and M5 have 
[Rb/Zr] ratios 0.4-0.5 dex higher (which may be due to systematic 
offsets). For M4, the difference arises due to 
$\omega$ Cen having less Rb and more Zr. For M5, the difference results from 
$\omega$ Cen having similar Rb and considerably more Zr. 

The similarity in heavy element abundance ratios involving Rb extends to
[Rb/La] - see middle panel of Figure \ref{fig:new3}. The indication
from our analyses is that the relative abundances of Rb to Y, Zr, and La are
the same for M5 and NGC 6752 as well as M4, 
the cluster clearly enriched 
in $s$-process elements. 
This demonstrates that the
source of the M4 enrichment is identical to that providing the $s$-process
elements for the natal material for M5 and NGC 6752. 
The abundance of Rb relative to the other elements is slightly
subsolar, i.e., [Rb/Y] $= -0.25$. There is a well known difference between the
spectroscopically determined Rb abundance and that obtained from carbonaceous
meteorites: $\log\epsilon$(Rb) = 2.60 versus 2.33, respectively 
\citep{asplund05}. If the meteoritic value is adopted [Rb/Y] and similar
ratios are increased by 0.27 dex and rendered indistinguishable
from the solar ratios. Then, $s$-process donors to the natal clouds of
the globular clusters and of local stars including the Sun must have
been similar. 

\subsection{Lead}

In M4 the mean Pb abundance is 
[Pb/Fe] = 0.30 $\pm$ 0.02 ($\sigma$ = 0.07) 
and in M5 the mean abundance is 
[Pb/Fe] = $-$0.35 $\pm$ 0.02 ($\sigma$ = 0.04). 
We note that 20\% of the solar Pb abundance is attributable to the 
$r$-process \citep{burris00} but the [Eu/Fe] ratio is comparable for 
M4 and M5. Again, since M4 has a lower [Eu/Fe] ratio than M5, the 
excess Pb in M4 relative to M5 cannot be due to an increased 
$r$-process contribution for M4. 
The excess likely arises 
from more complete processing down the $s$-process chain to its
termination at Pb. 
The difference in [Pb/Fe] between M4 and M5 is 0.65 dex. For 
the other neutron-capture elements synthesized via the $s$-process, 
the difference in [X/Fe] between M4 and M5 is typically 0.4 dex. 
Regarding the scatter in Pb abundances in M4 and M5, we again find that 
the small dispersion can be entirely explained by the measurement 
uncertainties. The Pb abundances in 
M4 and M5 are not correlated with O or Na. 

\begin{figure}
\epsscale{0.8}
\plotone{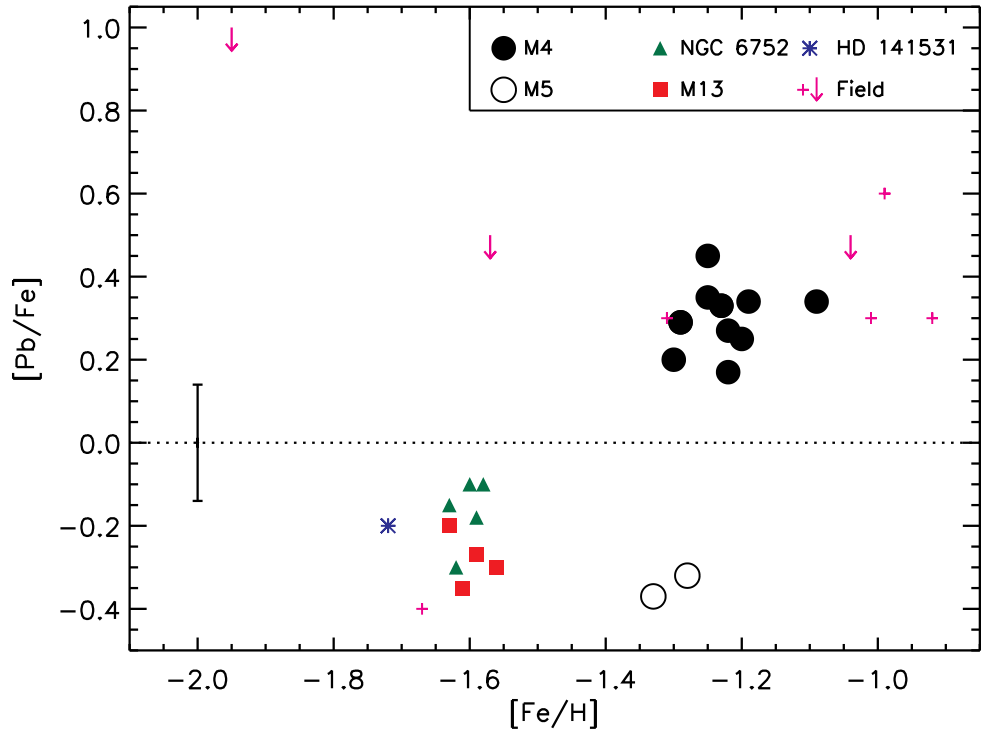}
\caption{[Pb/Fe] vs.\ [Fe/H] for M4 (filled circles) and M5 (open circles). 
Green triangles show NGC 6752, red 
squares represent M13, and blue asterisks are HD 141531 \citep{rbpbsubaru}. 
The magenta plus signs and upper limits represent field stars from 
\citet{sneden98} and \citet{travaglio01}. A representative error bar is 
shown.\label{fig:pb}}
\end{figure}

In Figure \ref{fig:pb}, we compare the [Pb/Fe] abundance ratios 
with field stars \citep{sneden98,travaglio01} as well as the globular clusters
M13 and NGC 6752 \citep{rbpbsubaru}.
Data on Pb in the field stars is obviously very limited. The one certain
data point is for HD 23439A, a star enriched in $s$-process products
(Tomkin \& Lambert 1999), with [Pb/Fe] $= +0.6$. The other three data points not
indicated as upper limits in
Figure \ref{fig:pb} are marked with the customary `:' by 
\citep{travaglio01} and inspection of their published spectra shows that
an alternative designation as an upper limit is possibly more
appropriate. Our Pb abundance determination for HD 141531 (Yong et al. 2006b)
is consistent with the sub-solar [Pb/Fe] ratios for NGC 6752, M5, and M13. 
Although more Pb abundances are needed for field stars, we suppose that
the Pb abundances of the globular clusters NGC 6752, M5, and M13 and 
normal field stars
are in fair agreement, as found for Rb, Y, Zr, Ba, and other heavy
elements.

In the lower panel of Figure \ref{fig:new3}, 
we plot [Rb/Pb] versus [Pb/Fe] for M4, M5, and NGC 6752. 
Interestingly, the [Rb/Pb] ratios for M4 and 
NGC 6752 appear similar which again suggests that the source of the 
$s$-process enrichment in M4 may be identical to the source of the 
$s$-process elements for NGC 6752. However, in this Figure, M5 now appears 
to be the outlier. 

Given the small comparison sample, clearly it would be of great interest 
to expand the measurements of Pb to larger samples of field stars. The 
synthetic spectra in this study and in \citet{rbpbsubaru} indicate that 
measurements of Pb in cool giants in the range 
$-$2.0 $\le$ [Fe/H] $\le$ $-$1.0 are feasible. 

\subsection{Consequences for the primordial scenario}

Formation of the Galactic and extragalactic globular clusters
remains a subject with many unanswered questions \citep{brodie06}. 
From the point of view of the quantitative
spectroscopist interested in the composition of stars in the
Galactic globular clusters, two questions seem preeminent: (i)
Why do the cluster and field stars follow the same
[X/Fe] versus [Fe/H] relations? (ii) What causes the light element
abundance variations first observed among cluster giant stars but now
found also in subgiant and main sequence stars? 

The simplest answer to question (i) would appear to be that
the clusters are formed during episodic mergers in the hierarchical
formation of the Galaxy, as discussed by \citet{bekki02} and 
\citet{beasley02}. If the gas
content of the protocluster is dominated by that from the
Galaxy and not the infalling system, the composition of the
cluster stars will be very similar to that of the Galactic
stars formed just preceding the merger. Gas remaining after star
formation in the cluster will be ejected by supernovae
from the massive stars. Star formation will be restricted to a
single generation of stars, i.e., stars within a given cluster
will have the identical metallicity. Indeed the
complete composition of all stars will be the same
 but for changes resulting from
stellar evolution and the agents responsible for the star-to-star
variations among light element abundances. These agents comprise
what is widely referred to as `the primordial scenario'.

For the majority of the Galactic globular clusters subjected to abundance
analysis, the compositions do closely resemble field stars of the
cluster's metallicity. This is as expected on the merger hypothesis.
M4 is an apparent exception with its overabundance of heavy
elements (Y, Zr, Ba, La etc, but not Eu) matched by very
few field stars. Some of the field stars with
 the cluster's degree of overabundance
of $s$-process products are binaries in which the products were
transferred to the present star when its companion was an AGB star.
There are very few known examples where mass transfer across a binary
is not a viable explanation. The case of HD 23439A and B \citep{tomkin99} 
would appear to be such a case. HD 23439A shows no sign of radial 
velocity variation during 14 years of monitoring \citep{latham02}. 
On the other hand, HD 23439B 
is a binary but its period and orbital eccentricity do not suggest 
mass transfer \citep{latham02}.

Other clusters
with few counterparts among field stars include Rup 106 and Pal 12. 
In their case, the Sagittarius dwarf spheroidal galaxy now undergoing
destruction by the Galaxy has giants with the peculiar composition of
Rup 106 and Pal 12 \citep{sbordone06} 
indicating that they belong to the dwarf spheroidal galaxy. 
A conjecture is that the
cluster M4 was captured from a stellar system in which
chemical evolution had led to $s$-process enrichment; a change in
the IMF might have been immediately responsible. 
However, consideration of space velocities reveals that M4's orbit 
is restricted to the inner disk and bulge (apocentric radius, $R_a$ = 
5.9 kpc) whereas M5's orbit may be more consistent with a capture 
event, $R_a$ = 35.4 kpc \citep{dinescu99}. Given M4's orbit, a 
comparison of the $s$-process elements in M4 and comparable 
metallicity stars of the inner disk and bulge would be of great interest. 

In the primordial scenario, the star-to-star abundance variations
apart from those attributable to dredge-up in giants arise from
the accretion of gas (or contamination of natal clouds) 
to varying degrees by cluster stars. 
The accreted gas is assumed to be mass lost from the more massive
and now dead stars of the cluster. Proposed candidates include 
IM-AGBs who experience extensive nucleosynthesis 
and lose their envelopes at low velocity so that 
the ejecta reside within the cluster's potential \citep{cottrell81}. 
Recently, a role for massive stars in polluting the
cluster's environment has been proposed 
\citep{decressin06,smith06,prantzos06}.

Both M4 and M5 exhibit star-to-star abundance variations for the light 
elements, as does every well studied Galactic globular cluster.
The variations involve O, Na, Mg, and Al, as well as C and N for
which there is a component related to dredge-up by giants.
For the majority of the analyzed clusters (M4 is an obvious exception),
the elemental abundances identified with the unpolluted
cluster stars match those of field stars of the same metallicity.
Two clusters have been studied for variations among the isotopic
ratios of Mg but
the normal stars in M13 and NGC 6752 (i.e., cluster stars with high O, high Mg, 
low Na, low Al and whose O-Al abundances match field stars at the same 
metallicity) had Mg isotope ratios that exceeded measurements in 
comparable metallicity 
field stars \citep{6752,mghdwarf,mghsubaru}.

In this study, we extended the elemental abundance measurements 
of Rb and Pb to the globular clusters M4 and M5. 
All four clusters that we have studied 
(M4, M5, M13, and NGC 6752) 
show light element variations, but 
constant Rb and Pb abundances within each cluster.
Thus, the accreted gas providing the light element variations 
cannot contain significant amounts (or significant deficiencies) 
of Rb and Pb. The accreted gas 
must have the same proportions of Rb and Pb as the ambient material. 

The source of the gas may be IM-AGBs that at least qualitatively
account for the range of abundance variations. 
A quantitative test is impossible at present
because the predicted yields from metal-poor IM-AGBs may be model dependent 
\citep{busso01,ventura05a,ventura05b,ventura05c}. 
As far as $s$-products are concerned which, as we have emphasised, do not
show abundance variations, 
 IM-AGBs remain a viable candidate because 
 \citet{lattanzio04} 
and Straniero (2006, private communication) suggest that IM-AGBs will not 
produce any $s$-process elements since the mass of the He shell and the 
duration of the thermal pulse decreases with increasing AGB mass. 

Massive stars achieve limited synthesis of $s$-process products but are held
to be the sites of the weak $s$-process (and the $r$-process)
 contributing nuclides
from the iron group up to about Rb, Y and Zr \citep{raiteri91}. Predicted
yields are such that massive stars could serve to provide the light
element abundance variations without leading to predicted variations
for Rb, Y, and Zr. 
 
In short, the lack of abundance variations for Rb, Y, Zr and
other s-process elements may not exclude either IM-AGBs or massive stars
as sources of the accreted pollutants.

\section{Concluding remarks}
\label{sec:summary}

In this paper we present measurements 
of the neutron-capture elements Rb and Pb 
in the globular clusters M4 and M5. While both clusters exhibit star-to-star
abundance variations for the light elements, we find that the 
abundances of Rb and Pb are constant. 
None of the abundance ratios [Rb/Fe], [Rb/Zr], and [Pb/Fe] are correlated 
with O or Na abundances.
In the primordial scenario, the abundance variations for the light
elements are attributed to different levels of accretion of
ejecta from IM-AGBs or massive stars. The fact that the heavy
elements including Rb and Pb do not show abundance variations implies
that the accreted material has the same composition as the ambient material 
for the heavy elements (i.e., the accreted material cannot be 
highly underabundant or overabundant in these elements). 
That the ratios [Rb/X] for X = Y, Zr, La are 
similar for M4 and M5 suggests that the source of the $s$-process 
elements are similar and that M4 had a greater concentration of 
these products. 

There remains a need to pursue additional observational tests of the
primordial scenario. In particular, present data on the Rb and Pb
abundances in field and cluster stars are sparse. The indication
that the Mg isotopic ratios of unpolluted or normal cluster
stars differ from those of field stars of the same metallicity
deserves closer scrutiny by, in particular, extending the
measurement of these isotopic ratios to additional clusters.

\acknowledgments

This research has made use of the SIMBAD database,
operated at CDS, Strasbourg, France and
NASA's Astrophysics Data System. 
We thank the anonymous referee for helpful comments and 
Inese Ivans for sending observed and synthetic spectra. 
DY thanks Amanda Karakas, Francesca D'Antona, 
Inese Ivans, John Lattanzio, Oscar Straniero, Paolo Ventura, 
and Roberto Gallino for helpful discussions and Chris Sneden for 
providing a line list for the Pb region. 
This research was performed while DBP held a National Research Council
Research Associateship Award at NASA's Goddard Space Flight Center.
DLL acknowledges support from the Robert A.\ Welch Foundation of Houston, 
Texas. BWC acknowledges support from the National Science Foundation 
through grant AST-0305431 to the University of North Carolina. 
This research was 
supported in part by NASA through the American Astronomical Society's Small 
Research Grant Program.

\clearpage

\clearpage

\clearpage

\clearpage

\clearpage

\clearpage

\begin{deluxetable}{lccccccc} 
\tabletypesize{\footnotesize}
\tablecolumns{8} 
\tablewidth{0pc} 
\tablecaption{Program stars, 
exposure times and stellar parameters.\label{tab:param}}
\tablehead{ 
\colhead{Star} &
\colhead{Exposure} &
\colhead{S/N\tablenotemark{a}} &
\colhead{S/N\tablenotemark{a}} &
\colhead{\teff} &
\colhead{log $g$} &
\colhead{$\xi_t$} &
\colhead{[Fe/H]} \\
\colhead{} &
\colhead{Time (min)} &
\colhead{4050\AA} &
\colhead{7800\AA} &
\colhead{K} &
\colhead{} &
\colhead{km s$^{-1}$} &
\colhead{}
}
\startdata
M4 L1411  & 40 & 61 & 397 & 4025 & 0.80 & 1.75 & $-$1.23 \\
M4 L1501  & 151 & 72 & 353 & 4175 & 1.00 & 1.55 & $-$1.29 \\
M4 L1514  & 35 & 59 & 432 & 3950 & 0.30 & 1.85 & $-$1.22 \\
M4 L2307  & 80 & 82 & 453 & 4125 & 0.95 & 1.65 & $-$1.19 \\
M4 L2406  & 30 & 72 & 395 & 4150 & 0.15 & 2.20 & $-$1.30 \\
M4 L2617  & 140 & 71 & 377 & 4275 & 1.25 & 1.65 & $-$1.20 \\
M4 L3209  & 35 & 68 & 407 & 4075 & 0.75 & 1.95 & $-$1.25 \\
M4 L3413  & 135 & 88 & 440 & 4225 & 1.10 & 1.75 & $-$1.23 \\
M4 L3624  & 80 & 39 & 285 & 4225 & 1.05 & 1.60 & $-$1.29 \\
M4 L4511  & 80 & 75 & 376 & 4150 & 1.05 & 1.70 & $-$1.22 \\
M4 L4611  & 60 & 29 & 393 & 3925 & 0.15 & 1.45 & $-$1.09 \\
M4 L4613  & 45 & 42 & 420 & 3900 & 0.20 & 1.70 & $-$1.25 \\
M5 IV-81  & 77 & 25 & 210 & 4050 & 0.30 & 1.90 & $-$1.28 \\
M5 IV-82  & 190 & 79 & 218 & 4400 & 1.20 & 1.75 & $-$1.33 \\
\enddata

\tablenotetext{a}{S/N values are per pixel.}

\end{deluxetable}

\begin{deluxetable}{lcrccrccrccr} 
\tabletypesize{\scriptsize}
\tablecolumns{12} 
\tablewidth{0pc} 
\tablecaption{Elemental abundances.\label{tab:abund}}
\tablehead{ 
\colhead{Star} &
\colhead{[O/Fe]} &
\colhead{[Na/Fe]} &
\colhead{[Mg/Fe]} &
\colhead{[Al/Fe]} &
\colhead{[Rb/Fe]} &
\colhead{[Y/Fe]} &
\colhead{[Zr/Fe]\tablenotemark{a}} &
\colhead{[Zr/Fe]\tablenotemark{b}} &
\colhead{[La/Fe]} &
\colhead{[Eu/Fe]} &
\colhead{[Pb/Fe]} 
}
\startdata
M4 L1411  & 0.46 & 0.55 & 0.55 & 0.80 & 0.45 & 0.74 & 0.62 & 0.32 & 0.58 & 0.49 & 0.33 \\
M4 L1501  & 0.57 & 0.51 & 0.59 & 0.79 & 0.37 & 0.69 & 0.43 & 0.13 & 0.48 & 0.47 & 0.29 \\
M4 L1514  & 0.69 & 0.10 & 0.50 & 0.63 & 0.32 & 0.71 & 0.62 & 0.32 & 0.28 & 0.25 & 0.17 \\
M4 L2307  & 0.47 & 0.50 & 0.54 & 0.76 & 0.44 & 0.79 & 0.53 & 0.23 & 0.49 & 0.46 & 0.34 \\
M4 L2406  & 0.51 & 0.46 & 0.58 & 0.66 & 0.25 & \ldots & \ldots & \ldots & 0.28 & 0.18 & 0.20 \\
M4 L2617  & 0.44 & 0.48 & 0.52 & 0.76 & 0.39 & 0.63 & 0.51 & 0.21 & 0.51 & 0.44 & 0.25 \\
M4 L3209  & 0.67 & 0.42 & 0.60 & 0.75 & 0.49 & 0.67 & 0.65 & 0.35 & 0.53 & 0.44 & 0.35 \\
M4 L3413  & 0.76 & 0.06 & 0.57 & 0.67 & 0.38 & 0.62 & 0.51 & 0.21 & 0.49 & 0.44 & 0.33 \\
M4 L3624  & 0.78 & 0.18 & 0.64 & 0.66 & 0.40 & 0.68 & 0.49 & 0.19 & 0.49 & 0.41 & 0.29 \\
M4 L4511  & 0.49 & 0.50 & 0.60 & 0.76 & 0.37 & 0.70 & 0.42 & 0.12 & 0.51 & 0.48 & 0.27 \\
M4 L4611  & 0.25 & 0.67 & 0.47 & 0.75 & 0.39 & \ldots & \ldots & \ldots & 0.51 & 0.29 & 0.34 \\
M4 L4613  & 0.59 & 0.67 & 0.62 & 0.86 & 0.47 & \ldots & \ldots & \ldots & 0.56 & 0.39 & 0.45 \\
M5 IV-81  & 0.50 & 0.15 & 0.35 & 0.33 & $-$0.04 & 0.22 & 0.25 & $-$0.05 & 0.17 & 0.46 & $-$0.32 \\
M5 IV-82  & 0.69 & $-$0.22 & 0.40 & 0.06 & 0.05 & 0.12 & 0.19 & $-$0.11 & 0.22 & 0.59 & $-$0.37 \\
\enddata

\tablenotetext{a}{Zr abundances when using the $gf$ values and solar abundance
adopted in \citet{67522}.}
\tablenotetext{b}{Zr abundances when shifted onto the \citet{smith00} scale 
-- see \citet{rbpbsubaru} for details.}

\end{deluxetable}

\begin{deluxetable}{lccrcccccccccccccc} 
\tabletypesize{\tiny}
\rotate
\tablecolumns{18} 
\tablewidth{0pc} 
\tablecaption{Line lists and equivalent 
widths for program stars.\label{tab:line}}
\tablehead{ 
\colhead{\AA} &
\colhead{Species} &
\colhead{$\chi$ (eV)} &
\colhead{$\log gf$} &
\colhead{L1411} &
\colhead{L1501} &
\colhead{L1514} &
\colhead{L2307} &
\colhead{L2406} &
\colhead{L2617} &
\colhead{L3209} &
\colhead{L3413} &
\colhead{L3624} &
\colhead{L4511} &
\colhead{L4611} &
\colhead{L4613} &
\colhead{IV-81} &
\colhead{IV-82} 
}
\startdata
6300.30 & 8.0 & 0.00 & $-$9.75 & \multicolumn{14}{c}{Abundances derived using synthetic spectra} \\
6363.78 & 8.0 & 0.02 & $-$10.25 & \multicolumn{14}{c}{Abundances derived using synthetic spectra} \\
4982.83 & 11.0 & 2.10 & $-$0.91 & \ldots & 93.4 & \ldots & 122.5 & \ldots & 92.2 & \ldots & \ldots & 69.5 & 104.7 & \ldots & \ldots & 74.4 & 29.7 \\
5682.65 & 11.0 & 2.10 & $-$0.71 & \ldots & 117.9 & 119.4 & \ldots & \ldots & 121.1 & \ldots & 96.1 & 96.8 & \ldots & \ldots & \ldots & 102.8 & 51.0 \\
5688.22 & 11.0 & 2.10 & $-$0.40 & \ldots & \ldots & \ldots & \ldots & \ldots & \ldots & \ldots & 114.7 & 114.7 & \ldots & \ldots & \ldots & \ldots & 69.6 \\
6154.23 & 11.0 & 2.10 & $-$1.56 & 69.7 & 52.4 & 43.4 & 60.6 & 44.2 & 50.2 & 53.1 & 26.4 & 30.2 & 57.0 & 88.0 & 79.7 & 36.6 & 8.9 \\
6160.75 & 11.0 & 2.10 & $-$1.26 & 95.2 & 75.0 & 66.5 & 85.5 & 69.7 & 75.0 & 79.0 & 47.0 & 49.6 & 82.4 & 112.8 & 104.4 & 57.4 & 19.7 \\
5711.09 & 12.0 & 4.35 & $-$1.73 & \ldots & \ldots & \ldots & \ldots & \ldots & \ldots & \ldots & \ldots & \ldots & \ldots & \ldots & \ldots & \ldots & 101.8 \\
6318.71 & 12.0 & 5.11 & $-$1.97 & \multicolumn{14}{c}{Abundances derived using synthetic spectra} \\
6319.24 & 12.0 & 5.11 & $-$2.20 & \multicolumn{14}{c}{Abundances derived using synthetic spectra} \\
5557.07 & 13.0 & 3.14 & $-$1.95 & 32.9 & 21.5 & 69.3 & 25.8 & 15.7 & 21.1 & 28.0 & 19.2 & 17.8 & 23.7 & 104.6 & 105.4 & 15.2 & 7.5 \\
6696.02 & 13.0 & 3.14 & $-$1.57 & \multicolumn{14}{c}{Abundances derived using synthetic spectra} \\
6698.67 & 13.0 & 3.14 & $-$1.89 & \multicolumn{14}{c}{Abundances derived using synthetic spectra} \\
4602.00 & 26.0 & 1.61 & $-$3.15 & \ldots & 121.3 & \ldots & \ldots & \ldots & \ldots & \ldots & \ldots & \ldots & \ldots & \ldots & \ldots & \ldots & 109.4 \\
4802.88 & 26.0 & 3.69 & $-$1.53 & 72.3 & 65.5 & 74.4 & 73.2 & 75.1 & 70.7 & 63.9 & 71.2 & \ldots & \ldots & 69.1 & 68.9 & 73.6 & 59.6 \\
4930.31 & 26.0 & 3.96 & $-$1.26 & 79.9 & 69.4 & \ldots & 78.0 & 75.3 & 72.2 & 86.2 & 73.1 & \ldots & \ldots & \ldots & \ldots & 78.6 & 66.4 \\
5054.64 & 26.0 & 3.64 & $-$1.94 & \ldots & 49.1 & \ldots & \ldots & 53.8 & \ldots & \ldots & \ldots & 52.5 & \ldots & \ldots & \ldots & 62.5 & 37.3 \\
5223.19 & 26.0 & 3.63 & $-$1.80 & 45.7 & 37.3 & \ldots & 42.5 & \ldots & 36.5 & 49.5 & 39.4 & \ldots & \ldots & \ldots & \ldots & 42.8 & \ldots \\
5242.49 & 26.0 & 3.63 & $-$0.98 & 107.6 & 99.2 & 111.4 & \ldots & \ldots & \ldots & 107.2 & 103.7 & 84.6 & 104.7 & 108.5 & 105.7 & 110.0 & 90.8 \\
5267.28 & 26.0 & 4.37 & $-$1.66 & 33.5 & \ldots & \ldots & \ldots & \ldots & \ldots & \ldots & \ldots & \ldots & \ldots & \ldots & \ldots & \ldots & \ldots \\
5288.53 & 26.0 & 3.69 & $-$1.53 & \ldots & \ldots & \ldots & \ldots & 87.1 & \ldots & \ldots & \ldots & \ldots & \ldots & \ldots & \ldots & \ldots & \ldots \\
5321.11 & 26.0 & 4.43 & $-$1.11 & 55.7 & \ldots & \ldots & 48.4 & 42.6 & 38.9 & \ldots & \ldots & 39.8 & \ldots & \ldots & \ldots & \ldots & 27.8 \\
5326.14 & 26.0 & 3.57 & $-$2.13 & 54.8 & 48.4 & 61.9 & \ldots & 53.8 & 49.3 & \ldots & \ldots & 50.4 & \ldots & \ldots & 54.3 & \ldots & \ldots \\
5364.86 & 26.0 & 4.44 & 0.21 & 114.5 & \ldots & \ldots & \ldots & \ldots & \ldots & \ldots & \ldots & 105.3 & \ldots & \ldots & \ldots & \ldots & \ldots \\
5365.40 & 26.0 & 3.57 & $-$1.04 & 102.1 & 89.5 & 108.7 & 98.7 & \ldots & \ldots & \ldots & \ldots & 90.3 & \ldots & \ldots & \ldots & \ldots & \ldots \\
5367.48 & 26.0 & 4.41 & 0.43 & 116.2 & \ldots & \ldots & \ldots & \ldots & \ldots & \ldots & \ldots & 110.1 & 107.7 & 119.8 & 110.6 & \ldots & \ldots \\
5379.57 & 26.0 & 3.69 & $-$1.53 & 81.6 & 68.7 & 87.3 & 77.7 & 83.7 & 75.3 & 82.2 & 75.5 & 71.6 & 76.8 & 83.2 & 83.2 & 84.2 & 63.0 \\
5412.80 & 26.0 & 4.43 & $-$1.78 & \ldots & \ldots & \ldots & \ldots & \ldots & \ldots & \ldots & \ldots & \ldots & \ldots & \ldots & \ldots & 27.6 & \ldots \\
5491.84 & 26.0 & 4.18 & $-$2.25 & \ldots & 13.4 & \ldots & 18.6 & 15.9 & 13.8 & \ldots & \ldots & 10.8 & 14.4 & 28.9 & \ldots & \ldots & \ldots \\
5525.54 & 26.0 & 4.23 & $-$1.15 & \ldots & 54.7 & 66.8 & \ldots & \ldots & \ldots & \ldots & 58.6 & 55.2 & 58.8 & \ldots & \ldots & \ldots & 41.1 \\
5618.63 & 26.0 & 4.21 & $-$1.29 & \ldots & 48.2 & \ldots & \ldots & 56.7 & 48.0 & \ldots & \ldots & \ldots & 54.7 & \ldots & \ldots & 55.2 & 40.4 \\
5661.35 & 26.0 & 4.28 & $-$1.82 & \ldots & \ldots & \ldots & \ldots & \ldots & 24.9 & \ldots & \ldots & \ldots & \ldots & \ldots & \ldots & \ldots & 15.0 \\
5705.47 & 26.0 & 4.30 & $-$1.42 & 43.2 & 35.6 & 44.9 & 41.4 & 38.6 & 38.9 & 43.5 & 38.7 & \ldots & 39.4 & 44.3 & 42.6 & 37.8 & 27.9 \\
5775.08 & 26.0 & 4.22 & $-$1.31 & 67.0 & 58.1 & 68.8 & 65.5 & 69.5 & 62.5 & 69.0 & 65.8 & \ldots & 63.8 & 62.2 & 65.4 & \ldots & 52.6 \\
5778.45 & 26.0 & 2.59 & $-$3.48 & \ldots & \ldots & \ldots & \ldots & \ldots & \ldots & \ldots & \ldots & \ldots & \ldots & \ldots & \ldots & 67.8 & 38.4 \\
5855.09 & 26.0 & 4.60 & $-$1.55 & 17.1 & 13.3 & \ldots & 18.3 & 13.8 & 14.7 & 20.1 & 16.7 & 14.9 & \ldots & 25.4 & 24.0 & 18.5 & \ldots \\
5909.97 & 26.0 & 3.21 & $-$2.64 & 79.7 & 62.5 & 85.1 & 73.7 & 71.6 & 63.9 & 80.3 & 68.1 & 60.6 & 69.3 & \ldots & 85.0 & 77.7 & 43.9 \\
5916.25 & 26.0 & 2.45 & $-$2.99 & \ldots & \ldots & \ldots & \ldots & \ldots & \ldots & \ldots & \ldots & 94.3 & \ldots & \ldots & \ldots & \ldots & 80.8 \\
6012.21 & 26.0 & 2.22 & $-$4.07 & 65.7 & 52.6 & 73.2 & 60.6 & 59.4 & 50.6 & 66.7 & 54.7 & 50.8 & 56.5 & 70.7 & 68.2 & 65.4 & 34.4 \\
6027.05 & 26.0 & 4.07 & $-$1.11 & 77.5 & \ldots & 84.0 & 78.6 & 84.7 & 72.7 & 79.4 & 74.6 & 73.0 & 76.5 & 81.7 & 76.9 & 78.8 & 60.9 \\
6120.24 & 26.0 & 0.91 & $-$5.97 & 66.9 & 43.3 & 82.9 & 56.6 & 55.5 & \ldots & 66.7 & 46.0 & 41.0 & 48.5 & 87.9 & 84.1 & 72.1 & 25.2 \\
6151.62 & 26.0 & 2.17 & $-$3.30 & 116.7 & 95.8 & \ldots & \ldots & \ldots & \ldots & \ldots & 97.4 & 94.7 & \ldots & \ldots & \ldots & \ldots & 83.0 \\
6165.36 & 26.0 & 4.14 & $-$1.49 & 52.3 & 44.3 & \ldots & 50.4 & 49.1 & 48.5 & 53.7 & 46.4 & 43.5 & 50.4 & \ldots & \ldots & \ldots & 33.4 \\
6180.20 & 26.0 & 2.73 & $-$2.64 & \ldots & 89.5 & \ldots & \ldots & \ldots & 90.2 & \ldots & \ldots & 88.0 & \ldots & \ldots & \ldots & \ldots & \ldots \\
6200.31 & 26.0 & 2.61 & $-$2.44 & \ldots & \ldots & \ldots & \ldots & \ldots & \ldots & \ldots & \ldots & 113.5 & \ldots & \ldots & \ldots & \ldots & \ldots \\
6229.23 & 26.0 & 2.84 & $-$2.85 & 79.2 & 65.6 & \ldots & 73.8 & 77.4 & 62.2 & \ldots & 68.0 & 63.3 & 70.4 & \ldots & \ldots & 81.8 & 51.6 \\
6232.64 & 26.0 & 3.65 & $-$1.28 & 111.3 & \ldots & \ldots & \ldots & \ldots & \ldots & 109.5 & 100.4 & 97.6 & 104.2 & \ldots & \ldots & 109.3 & 83.8 \\
6270.22 & 26.0 & 2.86 & $-$2.51 & 96.0 & 85.0 & 98.0 & 91.7 & 103.5 & 87.1 & 94.2 & 87.5 & 83.8 & 90.4 & 97.0 & 93.6 & 100.1 & 70.1 \\
6271.28 & 26.0 & 3.33 & $-$2.76 & 46.2 & \ldots & 51.7 & \ldots & 39.5 & 36.9 & 47.4 & 39.0 & 36.0 & 39.2 & 53.4 & 50.5 & 43.5 & 25.4 \\
6336.82 & 26.0 & 3.68 & $-$0.92 & \ldots & \ldots & \ldots & \ldots & \ldots & \ldots & \ldots & \ldots & 111.9 & \ldots & \ldots & \ldots & \ldots & \ldots \\
6353.84 & 26.0 & 0.91 & $-$6.48 & 34.1 & \ldots & 43.5 & \ldots & 28.2 & \ldots & 34.4 & 22.1 & 18.9 & \ldots & 49.8 & 45.2 & 35.7 & \ldots \\
6355.03 & 26.0 & 2.84 & $-$2.40 & \ldots & \ldots & \ldots & \ldots & \ldots & \ldots & \ldots & \ldots & 111.2 & \ldots & \ldots & \ldots & \ldots & 89.5 \\
6408.02 & 26.0 & 3.68 & $-$1.07 & \ldots & \ldots & \ldots & \ldots & \ldots & \ldots & \ldots & \ldots & 102.6 & \ldots & \ldots & \ldots & \ldots & 91.2 \\
6518.36 & 26.0 & 2.83 & $-$2.50 & \ldots & 88.3 & \ldots & \ldots & \ldots & \ldots & \ldots & \ldots & 87.7 & \ldots & \ldots & \ldots & 107.8 & \ldots \\
6574.23 & 26.0 & 0.99 & $-$5.00 & \ldots & 103.3 & \ldots & 119.5 & \ldots & \ldots & \ldots & 109.7 & 100.4 & 110.3 & \ldots & \ldots & \ldots & 82.6 \\
6575.02 & 26.0 & 2.59 & $-$2.73 & \ldots & \ldots & \ldots & \ldots & \ldots & \ldots & \ldots & 110.0 & 103.0 & 113.5 & \ldots & \ldots & \ldots & 87.1 \\
6581.21 & 26.0 & 1.48 & $-$4.71 & \ldots & \ldots & \ldots & \ldots & 99.2 & \ldots & \ldots & \ldots & \ldots & \ldots & \ldots & \ldots & \ldots & 40.0 \\
6609.11 & 26.0 & 2.56 & $-$2.69 & \ldots & \ldots & \ldots & \ldots & \ldots & \ldots & \ldots & \ldots & 103.4 & \ldots & \ldots & \ldots & \ldots & 94.2 \\
6625.02 & 26.0 & 1.01 & $-$5.37 & \ldots & \ldots & \ldots & \ldots & 114.5 & \ldots & \ldots & \ldots & \ldots & \ldots & \ldots & \ldots & \ldots & 61.1 \\
6648.08 & 26.0 & 1.01 & $-$5.92 & 70.7 & 54.8 & 79.9 & 65.5 & 66.4 & 52.3 & 71.4 & 47.9 & 42.7 & 60.1 & 86.6 & 81.3 & 70.8 & 25.5 \\
6699.16 & 26.0 & 4.59 & $-$2.17 & \ldots & 5.1 & 6.4 & \ldots & \ldots & 5.6 & 5.1 & \ldots & 4.7 & 4.9 & \ldots & 6.0 & 5.5 & \ldots \\
6739.52 & 26.0 & 1.56 & $-$4.82 & 72.8 & 53.2 & \ldots & \ldots & \ldots & 51.5 & 74.2 & 57.1 & 50.7 & 58.0 & \ldots & \ldots & 77.2 & \ldots \\
6793.25 & 26.0 & 4.07 & $-$2.39 & \ldots & \ldots & \ldots & \ldots & \ldots & \ldots & \ldots & \ldots & \ldots & \ldots & \ldots & \ldots & 17.0 & \ldots \\
6810.26 & 26.0 & 4.60 & $-$1.00 & 46.5 & 42.5 & 54.2 & 47.3 & 42.9 & 41.4 & 49.5 & 42.3 & 37.6 & 43.2 & 56.8 & 56.8 & 41.5 & 26.4 \\
6837.02 & 26.0 & 4.59 & $-$1.76 & \ldots & \ldots & 16.3 & \ldots & \ldots & \ldots & \ldots & 15.8 & \ldots & \ldots & \ldots & \ldots & 15.6 & 8.9 \\
6971.93 & 26.0 & 3.02 & $-$3.39 & 33.7 & \ldots & 39.6 & \ldots & \ldots & \ldots & 35.1 & 26.6 & 25.5 & \ldots & \ldots & \ldots & 32.0 & \ldots \\
7112.17 & 26.0 & 2.99 & $-$3.04 & 62.6 & 54.2 & \ldots & 61.9 & 61.7 & 55.1 & 63.0 & 52.0 & 50.5 & \ldots & \ldots & \ldots & 62.7 & 34.9 \\
7223.66 & 26.0 & 3.02 & $-$2.27 & 110.8 & 96.1 & \ldots & \ldots & \ldots & \ldots & 109.4 & 96.4 & 93.7 & \ldots & \ldots & \ldots & 110.8 & 79.9 \\
7401.69 & 26.0 & 4.18 & $-$1.66 & 45.0 & \ldots & 46.3 & 44.8 & \ldots & \ldots & 45.3 & \ldots & \ldots & \ldots & \ldots & \ldots & 45.4 & \ldots \\
7710.36 & 26.0 & 4.22 & $-$1.13 & 77.5 & 62.9 & 82.0 & 75.5 & 78.4 & 65.7 & 77.7 & 66.6 & 59.6 & 70.9 & 77.2 & \ldots & \ldots & \ldots \\
7723.20 & 26.0 & 2.28 & $-$3.62 & 97.6 & 79.4 & \ldots & 90.0 & 103.4 & 78.8 & 97.6 & 85.8 & 81.3 & 88.1 & 110.6 & 111.2 & 111.0 & 69.2 \\
7912.86 & 26.0 & 0.86 & $-$4.85 & \ldots & \ldots & \ldots & \ldots & \ldots & \ldots & \ldots & \ldots & \ldots & \ldots & \ldots & \ldots & \ldots & 111.4 \\
7941.09 & 26.0 & 3.27 & $-$2.33 & 67.7 & 57.4 & 77.0 & 63.5 & \ldots & 58.1 & 67.7 & 61.3 & 56.2 & 61.7 & 71.7 & 69.4 & 71.8 & 47.0 \\
8075.15 & 26.0 & 0.91 & $-$5.09 & \ldots & 118.6 & \ldots & \ldots & \ldots & \ldots & \ldots & \ldots & 114.7 & \ldots & \ldots & \ldots & \ldots & 91.2 \\
8204.10 & 26.0 & 0.91 & $-$6.05 & 77.4 & 46.4 & 94.8 & 62.4 & \ldots & 43.8 & 77.4 & 50.4 & 41.4 & 55.4 & 94.5 & 93.1 & 77.2 & 28.4 \\
4620.52 & 26.1 & 2.83 & $-$3.08 & 40.0 & 45.1 & 46.2 & 44.4 & 60.6 & \ldots & \ldots & \ldots & \ldots & 41.4 & \ldots & \ldots & \ldots & 54.0 \\
4993.36 & 26.1 & 2.81 & $-$3.49 & 44.6 & 36.6 & 48.8 & 40.4 & \ldots & \ldots & \ldots & \ldots & \ldots & \ldots & 53.5 & \ldots & 55.7 & 34.0 \\
5234.62 & 26.1 & 3.22 & $-$2.15 & \ldots & 74.5 & 78.8 & 74.8 & \ldots & 79.9 & \ldots & \ldots & \ldots & 72.8 & \ldots & 66.6 & \ldots & 84.4 \\
5325.55 & 26.1 & 3.22 & $-$3.22 & \ldots & \ldots & 34.2 & 37.0 & 60.5 & 39.9 & \ldots & \ldots & 33.9 & 35.0 & \ldots & 26.4 & \ldots & \ldots \\
5414.07 & 26.1 & 3.22 & $-$3.75 & \ldots & \ldots & \ldots & \ldots & 29.3 & \ldots & \ldots & \ldots & 18.0 & 17.1 & \ldots & \ldots & 20.0 & \ldots \\
5425.26 & 26.1 & 3.20 & $-$3.37 & 29.6 & 31.5 & 32.3 & 28.8 & 47.9 & 31.0 & 30.1 & 31.8 & 24.5 & 30.1 & 29.2 & 27.1 & 34.7 & 32.5 \\
5991.38 & 26.1 & 3.15 & $-$3.56 & \ldots & \ldots & \ldots & \ldots & \ldots & \ldots & \ldots & \ldots & \ldots & \ldots & \ldots & \ldots & \ldots & 25.1 \\
6084.11 & 26.1 & 3.20 & $-$3.81 & 12.8 & \ldots & 15.8 & 14.3 & 24.4 & 15.5 & 12.0 & 13.8 & 12.8 & 13.8 & \ldots & 9.9 & 20.3 & 17.9 \\
6149.26 & 26.1 & 3.89 & $-$2.72 & \ldots & 23.9 & \ldots & 24.1 & 37.7 & 23.9 & \ldots & \ldots & 16.9 & \ldots & \ldots & \ldots & 25.0 & 25.7 \\
6247.56 & 26.1 & 3.89 & $-$2.33 & 29.0 & 33.5 & 33.4 & 31.8 & 56.7 & 35.2 & 29.0 & 34.7 & 33.4 & 33.3 & 35.0 & 31.6 & 36.5 & 41.8 \\
6369.46 & 26.1 & 2.89 & $-$4.25 & 9.8 & 12.6 & 17.3 & 11.8 & 19.4 & 13.0 & 12.6 & 16.0 & 14.5 & 13.2 & 19.4 & \ldots & 17.1 & 17.7 \\
6416.92 & 26.1 & 3.89 & $-$2.74 & 23.5 & 26.2 & 29.6 & 26.5 & 37.0 & 24.2 & 24.6 & 25.7 & 25.6 & 26.3 & 31.2 & 29.2 & 29.6 & 27.3 \\
6432.68 & 26.1 & 2.89 & $-$3.71 & 29.8 & 33.9 & 35.0 & 33.2 & 50.8 & 33.4 & 33.4 & 34.6 & 33.7 & 34.4 & 31.7 & 29.6 & 39.6 & 39.5 \\
6456.38 & 26.1 & 3.90 & $-$2.08 & 36.6 & 41.9 & 41.7 & 40.6 & 74.4 & 43.1 & 36.4 & 42.3 & 41.9 & 42.0 & \ldots & 35.5 & 46.5 & 51.4 \\
6516.08 & 26.1 & 2.89 & $-$3.45 & \ldots & \ldots & 50.2 & 49.3 & 68.3 & \ldots & 47.8 & \ldots & \ldots & \ldots & 55.6 & 49.7 & 57.2 & 51.4 \\
7224.49 & 26.1 & 3.89 & $-$3.24 & \ldots & \ldots & \ldots & \ldots & \ldots & \ldots & \ldots & 9.2 & 9.0 & \ldots & \ldots & \ldots & 15.1 & 9.8 \\
7479.69 & 26.1 & 3.89 & $-$3.59 & \ldots & \ldots & 5.8 & 8.7 & \ldots & \ldots & 7.4 & 6.0 & 6.6 & \ldots & \ldots & \ldots & 7.9 & \ldots \\
7515.83 & 26.1 & 3.90 & $-$3.43 & \ldots & \ldots & \ldots & \ldots & \ldots & \ldots & \ldots & \ldots & \ldots & \ldots & \ldots & \ldots & \ldots & 5.5 \\
7711.72 & 26.1 & 3.90 & $-$2.54 & 22.3 & 26.3 & 25.5 & 24.7 & 42.1 & 29.1 & 23.3 & 26.2 & 28.2 & 28.1 & 25.6 & 25.5 & 29.0 & 31.6 \\
5200.42 & 39.1 & 0.99 & $-$0.57 & \ldots & 105.0 & \ldots & 121.5 & \ldots & 105.0 & \ldots & 111.9 & 102.1 & 116.0 & \ldots & \ldots & 112.8 & 72.5 \\
5509.91 & 39.1 & 0.99 & $-$1.01 & 110.0 & 97.0 & 116.0 & 109.0 & 114.0 & 96.0 & 113.0 & 100.0 & 99.0 & 101.0 & 120.0 & 114.0 & 90.0 & 68.0 \\
5544.61 & 39.1 & 1.74 & $-$1.08 & 39.0 & 28.0 & 52.0 & 32.0 & 31.0 & 29.0 & 38.0 & 28.0 & 31.0 & 32.0 & 72.0 & 71.0 & 20.0 & 12.0 \\
6127.44 & 40.0 & 0.15 & $-$1.06 & 107.0 & 58.0 & 126.0 & 80.0 & 47.0 & 55.0 & 105.0 & 63.0 & 56.0 & 68.0 & 145.0 & 143.0 & 76.0 & 15.0 \\
6134.55 & 40.0 & 0.00 & $-$1.28 & 106.0 & 53.0 & 126.0 & 78.0 & 44.0 & 49.0 & 104.0 & 58.0 & 49.0 & 63.0 & 146.0 & 143.0 & 76.0 & 14.0 \\
6143.20 & 40.0 & 0.07 & $-$1.10 & 109.0 & 57.0 & 128.0 & 84.0 & 49.0 & 55.0 & 108.0 & 63.0 & 55.0 & 69.0 & 149.0 & 147.0 & 78.0 & 14.0 \\
5303.53 & 57.1 & 0.32 & $-$1.35 & \multicolumn{14}{c}{Abundances derived using synthetic spectra} \\
6390.49 & 57.1 & 0.30 & $-$1.41 & \multicolumn{14}{c}{Abundances derived using synthetic spectra} \\
6645.13 & 63.1 & 1.37 & 0.20 & \multicolumn{14}{c}{Abundances derived using synthetic spectra} \\
\enddata

\end{deluxetable}

\begin{deluxetable}{lrrr} 
\tabletypesize{\footnotesize}
\tablecolumns{4} 
\tablewidth{0pc} 
\tablecaption{Abundance dependences on model parameters for
M4 L2307.\label{tab:parvar}}
\tablehead{ 
\colhead{Species} &
\colhead{\teff~+ 50} &
\colhead{$\log g$ + 0.2} &
\colhead{$\xi_t$ + 0.2}
}
\startdata
{\rm [Fe/H]} & $-$0.01 & 0.04 & $-$0.03 \\
{\rm [O/Fe]} & 0.01 & 0.02 & 0.02 \\
{\rm [Na/Fe]} & 0.06 & $-$0.07 & 0.01 \\
{\rm [Mg/Fe]} & 0.04 & $-$0.05 & 0.02 \\
{\rm [Al/Fe]} & 0.05 & $-$0.06 & 0.02 \\
{\rm [Rb/Fe]} & 0.08 & $-$0.05 & 0.03 \\
{\rm [Y/Fe]} & 0.00 & 0.03 & $-$0.02 \\
{\rm [Zr/Fe]} & 0.13 & $-$0.03 & 0.02 \\
{\rm [La/Fe]} & 0.02 & 0.02 & 0.01 \\
{\rm [Eu/Fe]} & $-$0.01 & 0.04 & 0.01 \\
{\rm [Pb/Fe]} & 0.10 & $-$0.09 & $-$0.02 \\
\enddata

\end{deluxetable}

\end{document}